\documentclass[11pt]{article}
\usepackage{amsmath, amsthm}
\usepackage{latexsym}
\usepackage{graphicx}
\makeatletter
\newfont{\footsc}{cmcsc10 at 8truept}
\newfont{\footbf}{cmbx10 at 8truept}
\newfont{\footrm}{cmr10 at 10truept}
\makeatother
\newtheorem{theorem}{\bf Theorem}

\newtheorem{lemma}{\bf Lemma}
\newtheorem{corollary}{\bf Corollary}

\begin{document}
\title{On an Inequality of Karlin and Rinott Concerning Weighted Sums of i.i.d.\ Random Variables}

\author{Yaming Yu\\
\small Department of Statistics\\[-0.8ex]
\small University of California\\[-0.8ex] 
\small Irvine, CA 92697, USA\\[-0.8ex]
\small \texttt{yamingy@uci.edu}}

\date{}
\maketitle

This note delivers an entropy comparison result concerning weighted sums of i.i.d.\ (independent and identically distributed) 
random variables.  The main result, Theorem \ref{main}, confirms a conjecture of Karlin and Rinott (1981). 

For a continuous random variable $X$ with density $f(x),\ x\in \mathbf{R}$, the (differential) entropy is defined as
$$H(X)=-\int f\log f,$$
and the more general $\alpha$-entropy, $\alpha>0$, is defined as
$$H_\alpha(X)=\frac{1}{1-\alpha} \log G_\alpha (X),$$
where
\begin{equation}
\label{alpha}
G_\alpha(X)=\int f^\alpha.
\end{equation}
It is convenient to define $H(X)=H_\alpha(X)=-\infty$ when $X$ is discrete, e.g., degenerate.  (Our notation differs from that 
of Karlin and Rinott 1981 here.)

We study the entropy of a weighted sum, $S=\sum_{i=1}^n a_i X_i$, of i.i.d.\ random variables $X_i$, assuming that the 
density $f$ of $X_i$ is {\it log-concave}, i.e., $supp(f)=\{x:\ f(x)>0\}$ is an interval and $\log f$ is a
concave function on $supp(f)$.  The main result is that $H(S)$ (or $H_\alpha(S)$ with $0<\alpha<1$) is smaller when the 
weights $a_1, \ldots, a_n$ are more ``uniform'' in the sense of {\it majorization}.  A real vector $\mathbf{b}=(b_1, \ldots, 
b_n)^\top$ is said to majorize $\mathbf{a}=(a_1, \ldots, a_n)^\top$, denoted $\mathbf{a}\prec \mathbf{b}$, if there exists a 
doubly stochastic matrix $T$, i.e., an $n\times n$ matrix $(t_{ij})$ where $t_{ij}\geq 0,\ \sum_i t_{ij}=1,\ j=1,\ldots, n,$ 
and $\sum_j t_{ij}=1,\ i=1, \ldots, n$, such that 
$$T\mathbf{b}=\mathbf{a}.$$
A function $\phi(\mathbf{a})$ symmetric in the coordinates of $\mathbf{a}=(a_1, \ldots, a_n)^\top$ is said to be {\it Schur 
convex}, if 
$$\mathbf{a}\prec \mathbf{b}\Longrightarrow \phi(\mathbf{a})\leq \phi(\mathbf{b}).$$
Basic properties and various applications of these two notions can be found in Hardy et al.\ (1964) and Marshall and Olkin 
(1979). 

\begin{theorem}
\label{main}
Let $X_1, \ldots, X_n$ be i.i.d.\ continuous random variables having a log-concave density on $\mathbf{R}$.  Then 
$H(\sum_{i=1}^n a_i X_i)$ is a Schur convex function of $(a_1, \ldots, a_n)\in \mathbf{R}^n$.  The same holds for 
$H_\alpha(\sum_{i=1}^n a_i X_i)$ if $0<\alpha<1$.
\end{theorem}

As an immediate consequence of Theorem \ref{main}, we have
\begin{corollary}
\label{coro}
In the setting of Theorem \ref{main}, subject to a fixed $\sum_{i=1}^n a_i$, the entropy $H(\sum_{i=1}^n a_i X_i)$ is
minimized when all $a_i$'s are equal.  The same holds if $H$ is replaced by $H_\alpha$ with $\alpha\in (0,1)$.
\end{corollary}

Note that Corollary \ref{coro} and hence Theorem \ref{main} need not hold without the assumption that the density of $X_i$ is
log-concave.  For example, if $X_i\sim {\rm Gam}(1/n, 1)$, i.e., a gamma distribution with shape parameter $1/n$, then the
equally weighted $\sum_{i=1}^n X_i,$ which has an exponential distribution, maximizes rather than minimizes the entropy $H$
among $\sum_{i=1}^n a_i X_i$ with $\sum_{i=1}^n a_i=n$.  For more entropy comparison results where log-concavity plays a role, 
see Yu (2009a, 2009b). 

Karlin and Rinott (1981) conjectured Theorem \ref{main} (their Remark 3.1, p.\ 110) and proved a special case (their Theorem 
3.1) assuming that i) $a_i>0$ and ii) $f(x)$, the density of the $X_i$'s, is supported on $[0,\infty),$ and admits a Laplace 
transform of the form 
$$\int_0^\infty e^{-sx} f(x)\, {\rm d}x =\left\{\prod_{i=1}^\infty (1+\beta_i s)^{\alpha_i}\right\}^{-1},$$
where $\alpha_i\geq 1,\ \beta_i\geq 0,$ and $0<\sum_{i=1}^\infty \alpha_i\beta_i<\infty$.  Their proof of this special case, 
however, is somewhat complicated and does not extend easily when the additional assumptions are relaxed.  A short 
proof of the general case is presented below. 

We shall make use of the {\it convex order} $\leq_{cx}$ between random variables.  For random variables $X$ and $Y$ on 
$\mathbf{R}$ with finite means, we say $X$ is smaller than $Y$ in the convex order, denoted $X\leq_{cx} Y$, if 
$$E\phi(X)\leq E\phi(Y),$$
for every convex function $\phi$.  Properties of $\leq_{cx}$ and many other stochastic orders can be found in Shaked and 
Shanthikumar (1994).

Lemma 1 relates the convex order $\leq_{cx}$ and log-concavity to entropy comparisons.  The basic idea is due to Karlin and 
Rinott (1981).  See Yu (2009b) for a discrete version that is used to compare the entropy between compound distributions 
on nonnegative integers.

\begin{lemma}
\label{lem1}
Let $X$ and $Y$ be continuous random variables on $\mathbf{R}$.  Assume $X\leq_{cx} Y$ and assume that the density of $Y$ is 
log-concave.  Then $H(X)\leq H(Y)$ and $H_{\alpha}(X)\leq H_{\alpha}(Y),\ 0<\alpha<1.$
\end{lemma}
{\bf Proof.}
Denote the density functions of $X$ and $Y$ by $f$ and $g$ respectively.  Note that because $g$ is log-concave, $E 
Y^2<\infty$, which implies $H(Y)<\infty$ as $H(Y)$ is bounded from above by the entropy of a normal variate with the same 
variance as $Y$.  Also, $X\leq_{cx} Y$ implies $EX^2\leq EY^2<\infty$, which gives $H(X)<\infty$.

Using $X\leq_{cx} Y$ and Jensen's inequality we obtain
\begin{align*}
H(Y)&=-\int g\log g\\
    &\geq -\int f\log g\\
    &\geq -\int f\log f\\
    &=H(X).
\end{align*}
All integrals are effectively over $supp(g)$ as $X\leq_{cx} Y$ implies that $f$ assigns zero mass outside of $supp(g)$ when 
$supp(g)$ is an interval.

To show $H_\alpha (Y)\geq H_\alpha(X)$, we can equivalently show $G_\alpha (Y)\geq G_\alpha (X)$, with $G_\alpha$ given by 
(\ref{alpha}).  From the log-concavity of $g$ and $\alpha<1$, it follows that $(\alpha-1)\log g$ and hence 
$g^{\alpha-1}=\exp[(\alpha-1)\log g]$ are convex.  We may use this and $X\leq_{cx} Y$ and H\"{o}lder's inequality to obtain
\begin{align*}
G_\alpha(Y)&=\left(\int gg^{\alpha-1}\right)^{\alpha}\left(\int g^\alpha\right)^{1-\alpha}\\
    &\geq \left(\int f g^{\alpha-1}\right)^{\alpha}\left(\int g^\alpha\right)^{1-\alpha}\\
    &\geq \int f^\alpha\\
    &=G_{\alpha}(X). \qed
\end{align*}

Lemma \ref{lem2} compares weighted sums of exchangeable random variables in the convex order.
\begin{lemma}
\label{lem2}
Let $X_i,\ i=1, \ldots, n,$ be exchangeable random variables with a finite mean.  Assume $(a_1, \ldots, a_n)\prec (b_1, 
\ldots, b_n),\ a_i, b_i\in \mathbf{R}$.  Then 
$$\sum_{i=1}^n a_i X_i\leq_{cx} \sum_{i=1}^n b_i X_i.$$
\end{lemma}

Theorem \ref{main} then follows from Lemmas \ref{lem1} and \ref{lem2} and the well-known fact that convolutions of 
log-concave densities are also log-concave.

{\bf Remark}
Lemma \ref{lem2} can be traced back to Marshall and Proschan (1965) (see also Eaton and Olshen 1972 and Bock et al.\ 1987).  
When $X_i$'s are i.i.d., Lemma \ref{lem2} is given by Arnold and Villase\~{n}or (1986) for the case $a_1=\ldots=a_n=1/n,\
b_1=0,\ b_2=\ldots =b_n=1/(n-1)$, and by O'Cinneide (1991) for $a_1=\ldots =a_n=1/n$.  Further 
discussions and generalizations of Lemma \ref{lem2} can be found in Ma (2000).  Some recent applications of Lemma \ref{lem2} 
in the context of wireless communications can be found in Jorswieck and Boche (2007).

\end{document}